\documentclass[a4paper,fleqn,usenatbib]{mnras}

\usepackage[T1]{fontenc}
\usepackage{ae,aecompl}

\usepackage{graphicx}	
\usepackage{amsmath}	
\usepackage{amssymb}	
\usepackage{tikz}

\usepackage{newtxtext,newtxmath}

\newcounter{ionstage}
\renewcommand{\ion}[2]{\setcounter{ionstage}{#2}%
  \ensuremath{\mathrm{#1\,\scriptstyle\Roman{ionstage}}}}

\newcommand\oi{[\ion{O}{1}]}
\newcommand\ha{\ensuremath{\mathrm{H\alpha}}}

\title[Observing external disc photoevaporation]{[O\,{\LARGE I}] 6300\AA\ emission as a probe of external photoevaporation of protoplanetary discs}

\author[Ballabio et al. ]
{\parbox{\textwidth}{Giulia Ballabio$^1$\thanks{E-mail: g.ballabio@qmul.ac.uk}, Thomas J. Haworth$^1$ and W. J. Henney$^2$
}\vspace{0.4cm}\\
\parbox{\textwidth}{$^{1}$ Astronomy Unit, School of Physics and Astronomy, Queen Mary University of London, London E1 4NS, UK \\
$^2$ Instituto de Radioastronom\'{i}a y Astrof\'{i}sica, Universidad Nacional Autónoma de M\'{e}xico, Apartado Postal 3-72, 58090 Morelia, Michoac\'{a}n, M\'{e}xico
}}

\begin{document}

\date{Accepted 2022 November 22. Received 2022 November 21; in original form 2022 September 30}

\pagerange{\pageref{firstpage}--\pageref{lastpage}} \pubyear{2020}

\maketitle
\label{firstpage}

\begin{abstract}
We study the utility of the \oi{} 6300\AA\, forbidden line for identifying and interpreting externally driven photoevaporative winds in different environments and at a range of distances. Thermally excited \oi{} 6300\AA\, is a well known tracer of inner disc winds, so any external contribution needs to be distinguishable.  In external winds, the line is not thermally excited and instead results from the dissociation of OH and we study how the line luminosity resulting from that process scales with the disc/environmental parameters.  We find that the line luminosity increases dramatically with FUV radiation field strength above around 5000\,G$_0$. The predicted luminosities from our models are consistent with measurements of the line luminosity of proplyds in the Orion Nebula Cluster. The high luminosity in strong UV environments alone may act as a diagnostic, but a rise in the [OI]-to-accretion luminosity ratio is predicted to better separate the two contributions. This could provide a means of identifying external photoevaporation in distant clusters where the proplyd morphology of evaporating discs cannot be spatially resolved.
\end{abstract}


\section{Introduction}
It is now widely established that planets are formed from circumstellar ``protoplanetary'' discs of material surrounding young stars \citep[e.g.][]{2018A&A...617A..44K, 2018ApJ...860L..13P, 2019Natur.574..378T, Armitage2018}. The evolution of these discs is mainly regulated by a combination of accretion and mass loss due to winds  \citep[e.g.,][]{2017RSOS....470114E, 2022arXiv220310068P, 2022arXiv220309821L}. These winds can be internal, originating from the inner disc and associated with the host star, or external and driven by external radiation fields.

Internal winds can be driven by irradiation and heating of the disc by the host star in a process called internal photoevaporation \citep{2009ApJ...690.1539G, 2010MNRAS.401.1415O, 2011MNRAS.412...13O, 2012MNRAS.422.1880O, 2015ApJ...804...29G, 2017ApJ...847...11W, 2018ApJ...857...57N, 2018ApJ...865...75N, 2019MNRAS.490.5596W, 2019MNRAS.487..691P, 2021MNRAS.508.3611P}. Internal winds can also be driven magnetohydrodynamically, through the magnetorotational instability \citep[MRI][]{1991ApJ...376..214B, 2009ApJ...691L..49S, 2016A&A...596A..74S} or magnetocentrifugal winds \citep{2007prpl.conf..277P, 2014prpl.conf..411T, 2016ApJ...818..152B, 2020ApJ...896..126G}. Characterising how these winds operate and their relative impacts on the disc evolution is a key goal at present, in particular because internal winds (particularly internal MHD winds) could be important for extracting angular momentum from the disc, which could facilitate accretion even if the disc is low turbulence/low viscosity \citep[e.g.][]{2021MNRAS.507.1106C, 2022MNRAS.512.2290T, 2022ApJ...926L..23H}.

Given the above there has been a dedicated effort to probe inner winds, to identify the driving mechanisms and characterise the mass loss and angular momentum extraction. However, a possible inner wind is only directly identified in a handful of cases to date \citep[e.g.][]{2013A&A...555A..73K, 2016Natur.540..406B, 2020A&A...640A..82T, 2021ApJS..257...16B, 2022arXiv220901969V}. Much more often the wind has to be identified through spatially unresolved spectral lines, such as the [NeII] 12.81\,${\mu}$m and [O\,\textsc{i}]  6300\AA\ forbidden lines of neon and oxygen \citep[e.g.][]{2008MNRAS.391L..64A, 2011ApJ...736...13P, 2014A&A...569A...5N, 2016ApJ...831..169S, 2018A&A...609A..87N, 2018ApJ...868...28F, 2019ApJ...870...76B, 2020MNRAS.496.2932B, 2020A&A...643A..32G, 2022MNRAS.517.3598W}.

External winds are significantly different to internal winds. To our knowledge they are driven only by irradiation, and are hence referred to as external photoevaporative winds \citep[for a recent review see][]{2022arXiv220611910W}. The sources of the external irradiation are other members of the star forming cluster, particularly OB stars which dominate UV production in a cluster \citep[e.g.][]{2015NewAR..68....1D}. External photoevaporation strips material from the outer disc, where there is a large mass reservoir that is least weakly bound to the central star \citep{1994ApJ...428..654H, 1998ApJ...499..758J, 2000ApJ...539..258R, 2004ApJ...611..360A, 2016MNRAS.457.3593F, 2018MNRAS.481..452H}. This can lead to large mass loss rates and hence rapid depletion and truncation of a disc, ultimately resulting in a lower disc lifetime \citep{1996AJ....111.1977M, 1999AJ....118.2350H,2016arXiv160501773G, 2018MNRAS.475.5460H, 2019MNRAS.490.5678C, 2020MNRAS.491..903W, 2020MNRAS.492.1279S, 2022MNRAS.512.3788Q}.

External photoevaporation is theoretically thought to be important for the evolution of a large fraction \citep{2020MNRAS.491..903W} of discs in massive stellar clusters, however this needs to be observationally verified
 \citep[note that the majority of discs studied at high resolution are relatively nearby, in sparse stellar groups with only weak ambient UV fields, e.g.][]{2015ApJ...808L...3A, 2018ApJ...869L..41A, 2020MNRAS.491..903W, 2021ApJS..257....1O}.
The easiest externally photoevaporating discs to detect are the ones in nearby, high UV environments such as the Orion Nebula Cluster (ONC), where the wind is so strong that the disc/wind takes on a spatially resolvable cometary appearance referred to as a proplyd \citep[e.g.][]{1998AJ....115..263O, 1999AJ....118.2350H, 2016ApJ...826L..15K, 2021MNRAS.501.3502H}.
Proplyds evolve very quickly due to the extremely high mass loss rates driven by external photoevaporation, with depletion timescales $M/\dot{M}$ that are typically $<0.1$\,Myr \citep{1999AJ....118.2350H}. This means that only a handful of the most recently introduced discs to the high UV environment are caught in the act of undergoing external photoevaporation at any given time \citep{2019MNRAS.490.5478W}. So although proplyds tell us that external photoevaporation definitely happens, and the timescale for external photoevaporation of proplyds is sufficiently short to tell us that proplyds must be continually being introduced into the high UV region,  the prevalence of external photoevaporation in massive stellar clusters is still difficult to precisely determine observationally. A key issue with the study of external photoevaporation in high UV environments is that proplyds cannot be spatially resolved in distant clusters. We therefore require the means of identifying and diagnosing external photoevaporation without relying on detecting a cometary morphology.

Another aspect to the challenge of observationally determining the importance of external photoevaporation is to verifying over what range of UV field strengths external photoevaporation is actually effective. Most proplyds are found in strong UV environments in close proximity (within $\sim0.1$\,pc) of O stars \citep{1999AJ....118.2350H, 2021MNRAS.501.3502H} where the UV field is in the range $10^5-10^7$G$_0$\footnote{G$_0$ is the unit of the Habing radiation field \citep{1968BAN....19..421H}. 1~G$_0$ is $1.6\times10^{-3}$\,erg\,s$^{-1}$\,cm$^{-2}$ over the wavelength range 912-2400\AA, which is the value that \cite{1968BAN....19..421H} measured in the solar neighbourhood.}. However \cite{2016ApJ...826L..15K} found proplyds in NGC 1977 in the vicinity of the B1 star 42 Ori, in a UV environment that was estimated to be $\sim3000$\,G$_0$. However the proplyd stars were all very low mass at $\leq0.4\,$M$_\odot$ \citep[meaning that the disc is easily unbound][]{2022MNRAS.512.2594H} and it is unknown if higher (solar) mass stars would also appear as proplyds in such an environment. Furthermore, as the UV field drops further, but is still predicted to lead to significant mass loss \citep{2018MNRAS.481..452H, 2020MNRAS.491..903W} even discs around low mass stars with winds may cease exhibiting a cometary appearance. How then, do we detect externally photoevaporating discs in common intermediate UV environments? To achieve this we require predictions and understanding of observational tracers.

\cite{2020MNRAS.492.5030H} explored the utility of CI for probing winds in intermediate UV environments based on models by \cite{2019MNRAS.485.3895H}. CI traces a layer outside of the CO dissociation front at a scale and speed where the wind can be both spectrally and spatially resolved with ALMA, resulting in unambiguous signatures of a wind that also yield diagnostics of the wind speed and temperature. This is yet to be observationally verified, however. An initial exploration with APEX was undertaken in August/September 2021, searching for CI towards the proplyds in NGC1977 but obtaining no detections. This was found to be because the stars there are low mass (most $\leq0.2$\,M$_\odot$) and their discs are in the final stages of clearing with only $\sim0.5$\,M$_{\textrm{jup}}$ discs and $\sim10\,$kyr remaining lifetimes \citep{2022arXiv220303928H}. The future of CI observations of winds therefore requires deeper observations of known evaporating discs (to test predictions based on the models and these non-detections), as well as searches for wind signatures at larger distances from the main UV sources in non-proplyds where the masses are higher.

Given the above challenges regarding discovering proplyds in distant regions and weaker UV environments, it is also prudent to explore as many other diagnostics and identifiers of external photoevaporation as possible. Here we aim to understand the observational characteristics of the oxygen [O\,\textsc{i}] 6300\AA\ line, which is observed in emission towards proplyds \citep{1998AJ....116..293B}. As we will discuss further below, \cite{1999ApJ...515..669S} explained observations of this line associated with ONC proplyds as a result of OH dissociation near the H-H$_2$ transition, but studied just a single system. Prior to this work, there are no predictions about how the amount of [OI] emission scales with the irradiation environment. Furthermore, for distant targets the [O\,\textsc{i}] line will be spatially unresolved, so we need to determine how to distinguish between the emission from internal and external winds \citep{2014A&A...569A...5N}.  Spatially unresolved [O\,\textsc{i}] line emission has already possibly identified an externally photoevaporating disc in the inner part of the $\sigma$ Ori cluster \citep{2009A&A...495L..13R, 2014A&A...569A...5N} which does suggest that spatially unresolved [O\,\textsc{i}] could have utility as a wider identifier of external photoevaporation. However first we need to determine and understand the variation of the line strength as a function of environment, and how it might be distinguished from internal winds. That is the goal of this paper.

\section{The origin of [O\,\textsc{i}]  6300\AA~emission in external photoevaporative winds from OH photodissociation}
\label{sec:origin}
In the case of internal photoevaporative winds [O\,\textsc{i}]  6300\AA~is produced by collisional excitation with, for example, hydrogen atoms or electrons \citep[e.g.][]{1995ApJ...452..736H, 2014A&A...569A...5N}. In the case of external photoevaporation the wind is driven by FUV radiation, and at only tens to a few thousand Kelvin is insufficicently hot to thermally excite the [O\,\textsc{i}]. However, the production of [O\,\textsc{i}]  6300\AA~emission can occur non-thermally as a result of OH photodissociation \citep[e.g.][introduced a simple model for this to predict the emission from Orion Nebula Cluster proplyds]{ 1998ApJ...502L..71S}. When this dissociation takes place some fraction of the resulting oxygen atoms are in an excitation state that leads to the production of [O\,\textsc{i}]  6300\AA~photons. The OH dissociation front is anticipated to be close to the H-H$_2$ transition (the molecular hydrogen dissociation front) since OH itself is formed by the reaction O + H$_2 \rightarrow$ OH + H and so won't reside further out in the wind. [O\,\textsc{i}]  6300\AA~emission is therefore  expected to trace the important H-H$_2$ transition part of the flow.

In order for [O\,\textsc{i}] emission to result from the dissociation of OH, OH needs to not have reacted out by some other pathway. We hence require that the dissociation rate be faster than other chemical reaction pathways that could remove OH. The key reactions, equations that characterise their rates and relevant rate constants are summarised in Table \ref{tab:reactions}. Included are the rate coefficients from \cite{1985ApJ...291..722T}, used by \cite{1998ApJ...502L..71S}, as well as the UMIST rate coefficients \citep{2013A&A...550A..36M}, which are used by the \textsc{torus-3dpdr} code employed in this paper \citep[see section \ref{sec:postprocessing} and][]{2015MNRAS.454.2828B, 2019A&C....27...63H}.

\begin{table*}
    \centering
    \begin{tabular}{cccccccccc}
    \hline
    \hline
    \multicolumn{1}{c}{Reaction} & \multicolumn{3}{c}{UMIST network} & \multicolumn{3}{c}{\cite{1985ApJ...291..722T}} & $T_{min}$ \textrm{[K]} & $T_{max}$ \textrm{[K]} & Rate \\
    coefficients     & $a$ & $b$ & $c$ & $a$ & $b$ & $c$ & &  & equation \\
    \hline
    \hline
    \multicolumn{10}{c}{\textbf{OH FORMATION}} \\
    $\textrm{O} + \textrm{H}_2 \rightarrow \textrm{OH} + \textrm{H}$ & $3.14\times10^{-13}$ & 2.7 & 3150 & $9\times10^{-12}$ & 1.0 & 4500 & -- & -- & $a n^2X(\textrm{O})X(\textrm{H}_2)\left(\frac{T}{300}\right)^b
\exp \left( — \frac{c}{T}\right)$ \\
    \hline
    \multicolumn{10}{c}{\textbf{OH DESTRUCTION}} \\
    \multicolumn{10}{c}{Photodissociation} \\ 
    $\textrm{OH} + \nu \rightarrow \textrm{O} + \textrm{H}$ & $3.5\times10^{-10}$ & -- & 1.7 &  $2.2\times10^{-10}$ & -- & 2 & -- & -- & $a nX(\textrm{OH}) F_{\textrm{FUV}}\exp\left(-c A_V\right)$ \\
    \multicolumn{10}{c}{Other reactions} \\ 
    $\textrm{OH} + \textrm{C}^+ \rightarrow \textrm{CO}^+ + \textrm{H}$& $7.7\times10^{-10}$& -- & -- & $7.7\times10^{-10}$ & -- & -- & 10 &41000 &  $a n^2X(\textrm{OH})X(\textrm{C}^+)\left(\frac{T}{300}\right)^{1.25}$ \\
    $\textrm{OH} + \textrm{H} \rightarrow \textrm{O} + \textrm{H}_2$ & $6.99\times10^{-14}$ & 2.8 & 1950. & $4.2\times10^{-12}$ & 1.0 & 3500 & 300 & 2500 &  $a n^2X(\textrm{OH})X(\textrm{H}) \left(\frac{T}{300}\right)^b
\exp \left( — \frac{c}{T}\right)$ \\
    $\textrm{OH} + \textrm{H}_2 \rightarrow \textrm{H}_2\textrm{O} + \textrm{H}$& $2.05\times10^{-12}$ & 1.52 & 1736. & $3.6\times10^{-11}$ & 1.0 & 2600 & -- & -- & $a n^2X(\textrm{OH})X(\textrm{H}_2)\left(\frac{T}{300}\right)^b
\exp \left( — \frac{c}{T}\right)$  \\
    $\textrm{OH} + \textrm{O} \rightarrow \textrm{H} + \textrm{O}_2$& $4.3\times10^{-11}$ & -0.5 & 30.0 & $4\times10^{-10}$ & 1.0 & 600 & -- & -- &  $a n^2X(\textrm{OH})X(\textrm{O})\left(\frac{T}{300}\right)^b
\exp \left( — \frac{c}{T}\right)$  \\
    $\textrm{OH} + \textrm{OH} \rightarrow \textrm{H}_2\textrm{O} + \textrm{O}$ & $1.65\times10^{-12}$ & 1.14 & 50.0 & $10^{-11}$& 1.0 &550 & -- & -- &  $a n^2X(\textrm{OH})X(\textrm{OH})\left(\frac{T}{300}\right)^b
\exp \left( — \frac{c}{T}\right)$  \\
    \hline
    \hline
    \end{tabular}
    \caption{A summary of the key reactions that result in the the removal of OH. The top-most reaction is the photodissociation reaction that leads to the [O\,\textsc{i}]  emission that is the focus of this paper. The other reactions would remove OH by other means and hence suppress [O\,\textsc{i}]  emission. The code used for disc photoevaporation models in this paper, \textsc{torus-3dpdr}, is using a reduced version of the UMIST network (with coefficients in columns 2-4). In columns 5-7 we include the rate coefficients from \protect\cite{1985ApJ...291..722T}, which were used in \protect\cite{1998ApJ...502L..71S}. $T_{min}$ and $T_{max}$ denote temperature constraints on whether the reaction can take place, and the final column shows the rate equation, which the prior coefficients feed into.  }
    \label{tab:reactions}
\end{table*}

The photodissociation rate per unit volume is parameterised by
\begin{equation}
    R_{phot} = n X(\textrm{OH}) aF_{\textrm{FUV}}\exp\left(-c A_V\right) \textrm{cm}^{-3}\,\textrm{s}^{-1}
    \label{equn:photodiss}
\end{equation}
which intuitively falls off exponentially with increasing extinction $A_V$, and where $n$ is the number density,  $X(\textrm{OH})$ is the abundance of OH, $F_{\textrm{FUV}}$ is the FUV radiation field strength in Habing units and $a$ and $c$ are process-specific constants.

Similarly, for the reaction rate per unit volume for some 2-species (A and B) reaction in a medium of number density $n$ and temperature $T$ is
\begin{equation}
    R_{reac} = n^2 X(\textrm{A}) X(\textrm{B}) a\left(\frac{T}{300}\right)^b
\exp \left( — \frac{c}{T}\right)\,\textrm{cm}^{-3}\textrm{s}^{-1}
\end{equation}
which contains a third rate constant $b$.

In any given portion of the disc/wind we therefore require $R_{phot} > R_{reac}$ for OH to be dissociated more quickly than it is reacted out and hence to lead to the production of [O\,\textsc{i}]  6300\AA\, emission. In their work, trying to understand [O\,\textsc{i}] emission from proplyds, \cite{1998ApJ...502L..71S} approached this considering reactions of OH with C$^+$, H and H$_2$, and by assuming that in the wind near the H-H$_2$ transition the hydrogen is all atomic and that the abundance of C$^+$ is $3\times10^{-4}$. Applying rate coefficients from \cite{1985ApJ...291..722T} permitted them to derive three criteria that must be satisfied for OH removal by photodissociation and hence [O\,\textsc{i}]  emission to result
\begin{itemize}
    \item $F_{\textrm{FUV}} \geq 10^{-3} n$,
    \item $F_{\textrm{FUV}} \geq 10^{-4} T\exp(-3500/T)n$,
    \item $n < n_{\rm cr} \simeq 7\times 10^9$\,cm$^{-3}$. 
\end{itemize}
In this paper we will be postprocessing grid-based photodissociation-dynamics simulations which include the density and temperature structure in addition to abundances (detailed further in section \ref{sec:postprocessing}). We therefore will not employ these \cite{1998ApJ...502L..71S} constraints, and rather permit [O\,\textsc{i}]  emission on a cell-by-cell basis wherever the OH dissociation rate is faster than any other chemical pathways according to the UMIST coefficients and expressions in table \ref{tab:reactions}. 

We compared the photodissociation and reaction rates as a function of extinction and temperature respectively for the  \cite{1985ApJ...291..722T} \citep[used by][]{1998ApJ...502L..71S} and more recent UMIST coefficients (used here). The photodissociation rates are very similar. The reaction with the fastest rate is with C$^+$ (assuming C$^+$ is abundant), which has identical rates in \cite{1985ApJ...291..722T} and UMIST. The main differences arise in the reactions between OH and O or another OH. However, these are not dominant at the $\sim30-3000\,K$ anticipated for the FUV heated wind. We therefore do not expect major differences between our work and \cite{1998ApJ...502L..71S} based on updated rate constants.

\section{Theoretical estimates of the [O\,I] emission}
\label{sec:postprocessing}

\subsection{Numerical method}
\subsubsection{Radiation hydrodynamic models}
External photoevaporative winds are predominantly driven by FUV irradiation, which heats the disc/wind through photodissociation region (PDR) microphysics. Solving this numerically is computationally expensive due to the need to solve 3D line cooling. 1D models have provided a way to explore external photoevaporation in different scenarios \citep{2004ApJ...611..360A, 2016MNRAS.457.3593F, 2018MNRAS.481..452H} and have been demonstrated to do a good job of estimating mass loss rates compared to vastly more expensive 2D models \citep{2019MNRAS.485.3895H}. We therefore take an initial exploration of the anticipated scaling of the [O\,\textsc{i}]  6300\AA\ emission as a function of star/disc/UV parameters using 1D models.

The 1D models that we run are very similar to those in the FRIED grid of mass loss rates \cite{2018MNRAS.481..452H} and so we only give a brief overview here. In these models the steady state wind structure is solved for when a fixed disc boundary condition is irradiated by a fixed external UV source. The opacity in the wind accounts for the fact that only small dust grains are entrained in the wind and that therefore the dust-to-gas ratio is low in the wind if grain growth has occurred \citep{2016MNRAS.457.3593F}. We run our own models rather than utilise the underlying models from FRIED so that we can explore a bespoke parameter space (e.g. up to higher UV fields). We also explore the impact of varying the Polycyclic Aromatic Hydrocarbons (PAHs) abundance in our models, since photoelectric heating by PAHs is the dominant heating mechanism in the FUV-irradiated PDR of an external wind, but have a poorly constrained abundance (in FRIED the PAH-to-dust mass ratio was assumed to be one tenth of that in the interstellar medium). Note that only small grains are entrained in the photoevaporative wind, so if grain growth has occurred in the disc \citep[which happens fast, ][]{2018ApJ...869L..45B} the dust abundance in external photoevaporative winds is expected to be depleted \citep{2016MNRAS.457.3593F, 2021MNRAS.508.2493O}. Motivated by \cite{2016MNRAS.457.3593F} these models therefore have a dust-to-gas mass ratio of $3\times10^{-4}$.

\subsubsection{Estimating the [O\,\textsc{i}]  6300\AA\ line luminosity due to external photoevaporation}
Once the flow structure in the 1D dynamical models is solved for, we estimate the [O\,\textsc{i}] 6300\AA\ line intensity contribution from external wind as follows. Note that we are not including any contribution from internal winds and will discuss that further in \ref{sec:scaling} and \ref{sec:LOIvsLacc}.

We integrate from the disc outer edge to the outer extent of the simulation domain. In any given cell we first determine whether the local OH will be photodissociated or instead react out by some other pathway using the criteria discussed in section \ref{sec:origin}. That is, we require the photodissociation rate (equation \ref{equn:photodiss}) to be greater than the other possible reaction rates. If this is not the case then that cell's contribution to the line intensity is zero. Otherwise, the emission from cell $i$ of width $\Delta R$ is added to the total line intensity as
\begin{equation}
    I_{i} = I_{i-1} + j_\nu \Delta R
\end{equation}
where $j_\nu$ is the line emission coefficient, integrated over the entire line
\begin{equation}
    j_{\nu} = \frac{1}{4\pi}h\nu_0 n_u A_{ul}
\end{equation}
where $\nu_0$ is the rest frequency of the line, $n_u$ is the number density of particles in the upper excited state of the transition (and hence available to de-excite and emit), $A_{ul}$ is the Einstein A coefficient for the transition \citep[obtained from the LAMDA database][]{2005A&A...432..369S}. Line absorption is not a significant effect for forbidden lines (one of the characteristics that makes them efficient cooling mechanisms) because of the low absorption cross section (due to the low oscillator strength of the transition), so we do not undertake full velocity dependent line transfer. Since only small grains are entrained \citep{2016MNRAS.457.3593F} the dust-to-gas mass ratio in the external photoevaporative wind is also highly depleted (easily by at around 2 orders of magnitude), especially if grain growth and radial drift has occurred in the disc. We therefore also neglect the dust opacity, though including silicate grain opacities at the expected abundances makes a negligible difference to the emergent intensity.

The number of particles in the upper state of the transition $n_u$ is estimated using a similar approach to \cite{1998ApJ...502L..71S}, which assumes
  equilibrium between dissociation pumping and radiative decay. The state is populated when OH molecules are photodissociated, so it is not viable to take an approach such as using the Boltzmann distribution. The number density of particles in the upper state in a given cell is then
\begin{equation}
    n_u = \frac{1}{A_{ul}}n(\textrm{OH})R_{\textrm{OH},0}F_{\textrm{FUV}} f
    \label{equn:nupper}
\end{equation}
where $n(\textrm{OH})$ is the number density of OH molecules, $R_{\textrm{OH},0}$ the photodissociation rate of OH under an FUV radiation field of 1\,G$_0$, $F_{\textrm{FUV}}$ the FUV field strength in the cell in units of G$_0$ and $f$ the fraction of OH photodissociation events that will go on to result in the emission of a [O\,\textsc{i}]  6300\AA\ photon (we assume $f=0.5$).

The above integration determines the 1D line intensity $I_\nu$. Which is converted into a line luminosity via
\begin{equation}
    L_{\textrm{[OI]}} = 4\pi^2 R_d^2 I_\nu
\end{equation}
where $R_d$ is the disc radius. That is, we are assuming that the emission comes from a surface of order the disc size.

We wish to emphasise that here we are calculating the expected contribution of the external photoevaporative wind to the [OI] 6300\AA~line luminosity. Additional contributions from internal winds/magnetospheric accretion or outflows/jets are ignored (the latter of which can be distinguished in practice based on the high velocity). This means that our models would likely be lower limits on the actual [OI] luminosity, especially in low UV environments.

\subsection{Scaling of the line luminosity}
\label{sec:scaling}
We begin by using the large grid of 1D external photoevaporation models to study the general scaling of the \oi{} intensity/flux as a function of the star/disc/UV field parameters.

\begin{figure}
    \centering
    \includegraphics[width=\columnwidth]{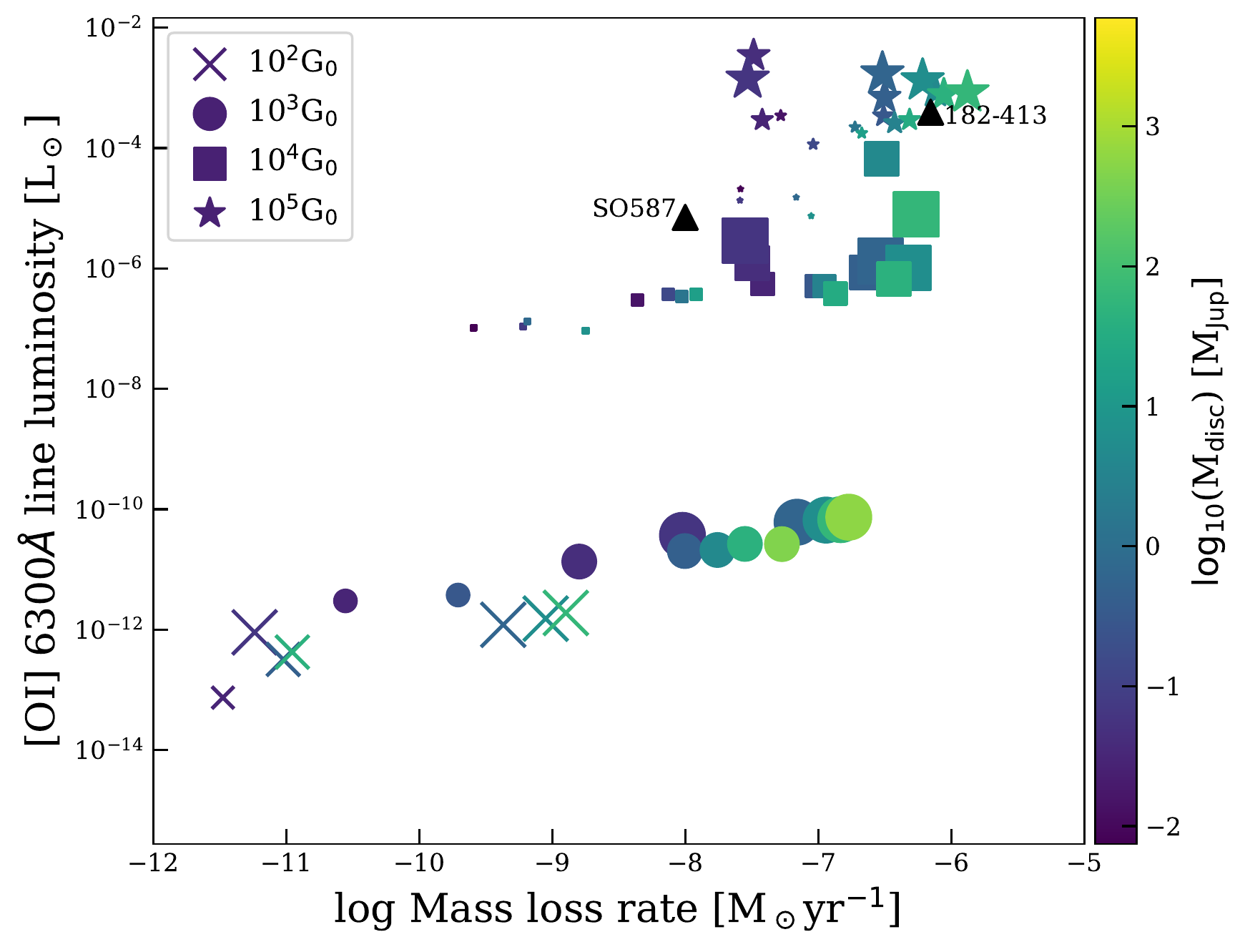}
    \caption{Predicted 1D model \oi{} 6300\AA\ line luminosities in the case of a 0.6\,$M_\odot$ star. The mass loss rates are due to external photoevaporation.  Crosses, circles, squares and stars correspond to UV fields of $10^2$, $10^3$, $10^4$ and $10^5$G$_0$ respectively. The symbol sizes scale with the disc radius, including 10, 20, 40, 60, and 80\,au. The colour scale represents the disc mass. By far the dominant factor setting the line luminosity is the UV field strength. The black triangles are the candidate evaporating system SO587 in $\sigma$ Ori \protect\citep{2009A&A...495L..13R} and the ONC proplyd 182-413 \protect\citep{1998ApJ...502L..71S, 1998AJ....116..293B}.}
    \label{fig:OILuminosities}
\end{figure}

\begin{figure*}
    \centering
    \includegraphics[width=1.6\columnwidth]{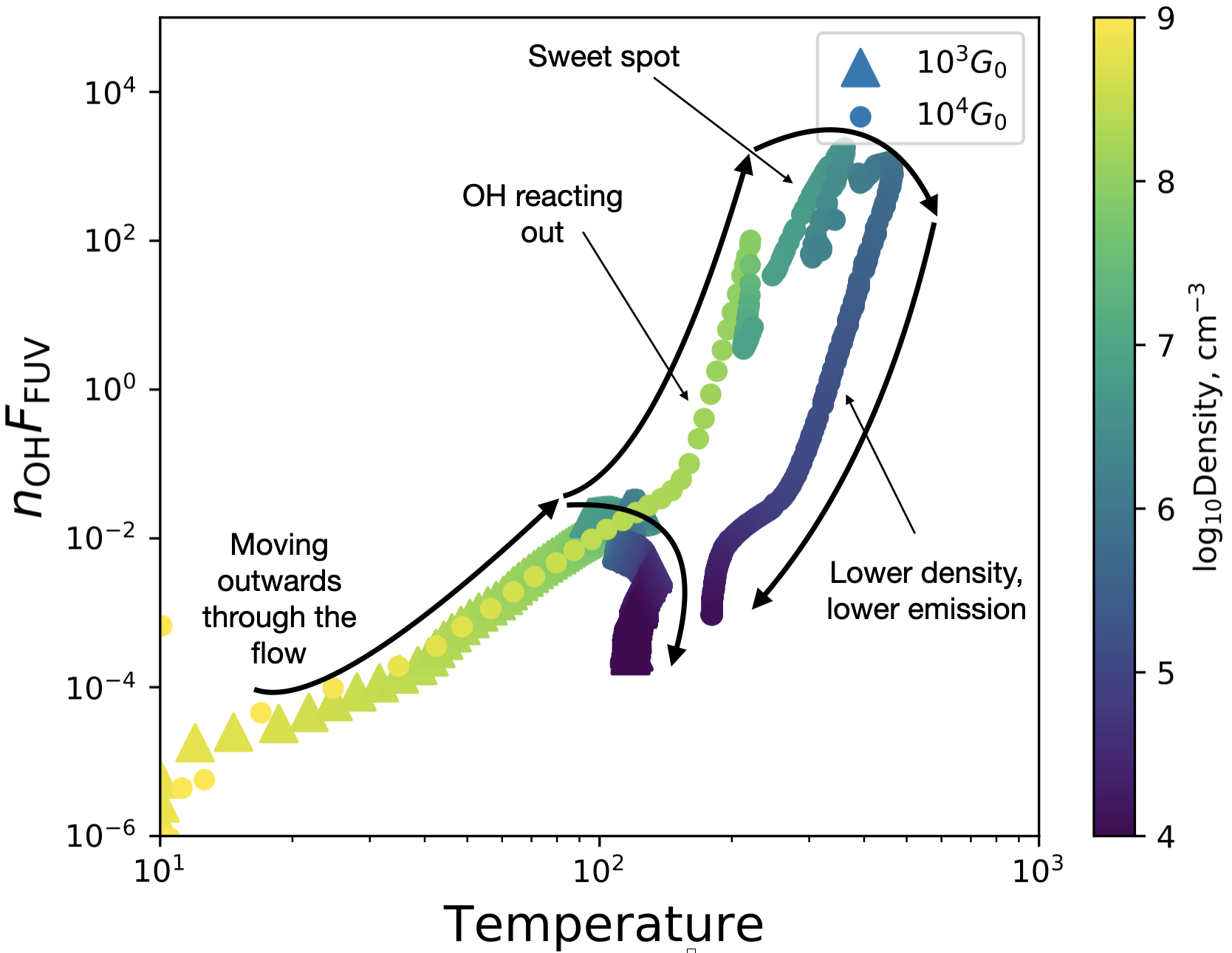}
    \caption{The parameter that controls the \oi{} emission measure as a function of the local temperature, with the density represented by the colour scale. The large triangles are a $10^3$G$_0$ environment and circles $10^4$G$_0$ environment. Denser parts of the flow at temperatures above 200\,K are key to bright \oi{} emission, which are only reached beyond $10^3$G$_0$. This is a comparison of a 0.6\,M$_\odot$ star, with $\Sigma_{1au}=10$\,g\,cm$^{-2}$ and a disc radius of 80\,au.}
    \label{fig:EmissionMeasure}
\end{figure*}

Figure \ref{fig:OILuminosities} shows the line luminosity for a grid of models around 0.6\,M$_\odot$ stars. The crosses, circles, square and star symbols correspond to discs irradiated by $10^2, 10^3, 10^4$ and $10^5$G$_0$ UV fields respectively. The colour scale represents different disc masses and the symbol sizes scale with the disc radius (included are models of radius 10, 20, 40, 60 and 80\,au). The main implication of Figure \ref{fig:OILuminosities} is that the \oi{} 6300\AA\ line luminosity is predominantly sensitive to the external UV field strength and very insensitive to the disc mass and radius. Furthermore the line intensity undergoes a massive increase of four orders of magnitude from $10^3$\,G$_0$ to $10^4$\,G$_0$.

\begin{figure*}
    \centering
    \includegraphics[width=\textwidth]{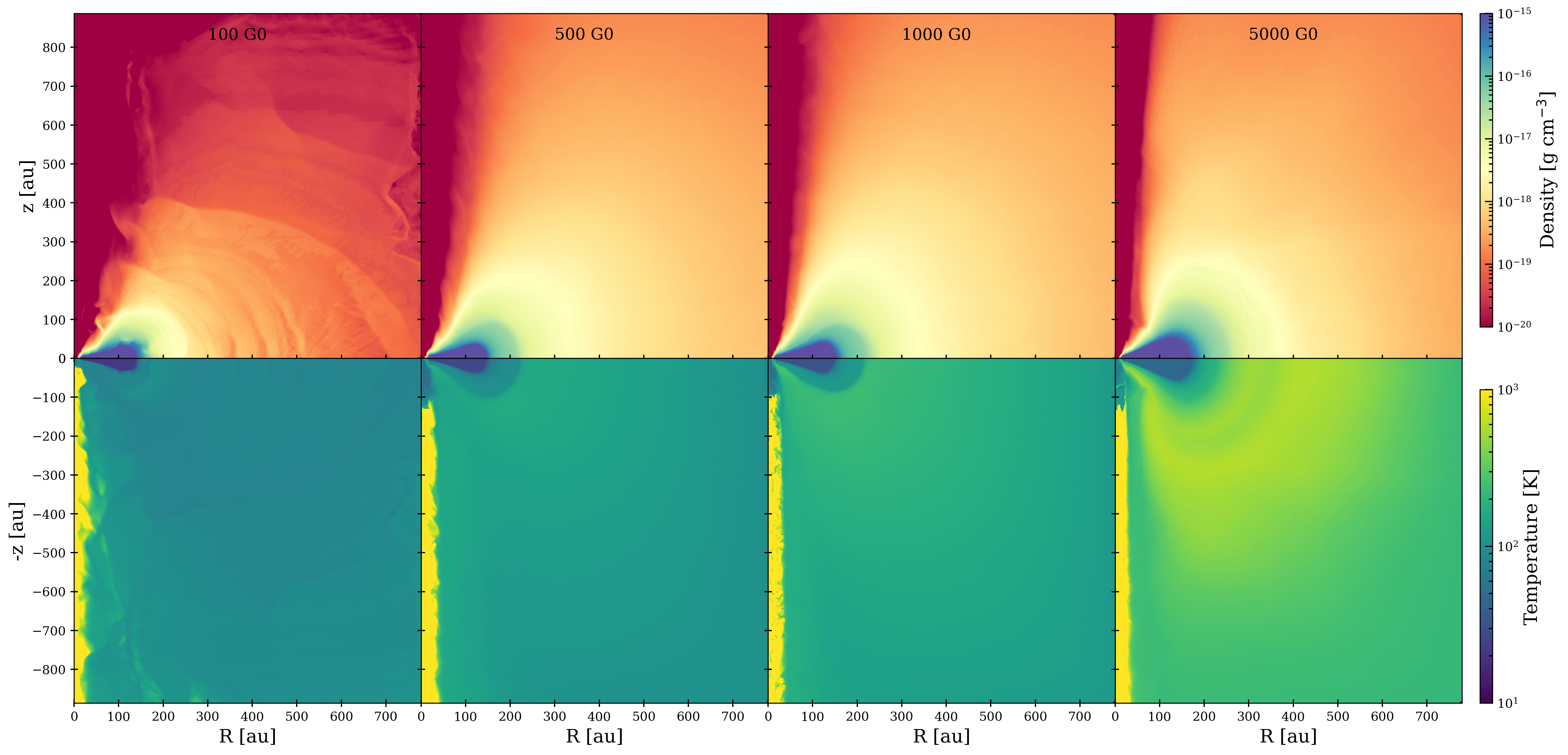}
    \includegraphics[width=\textwidth]{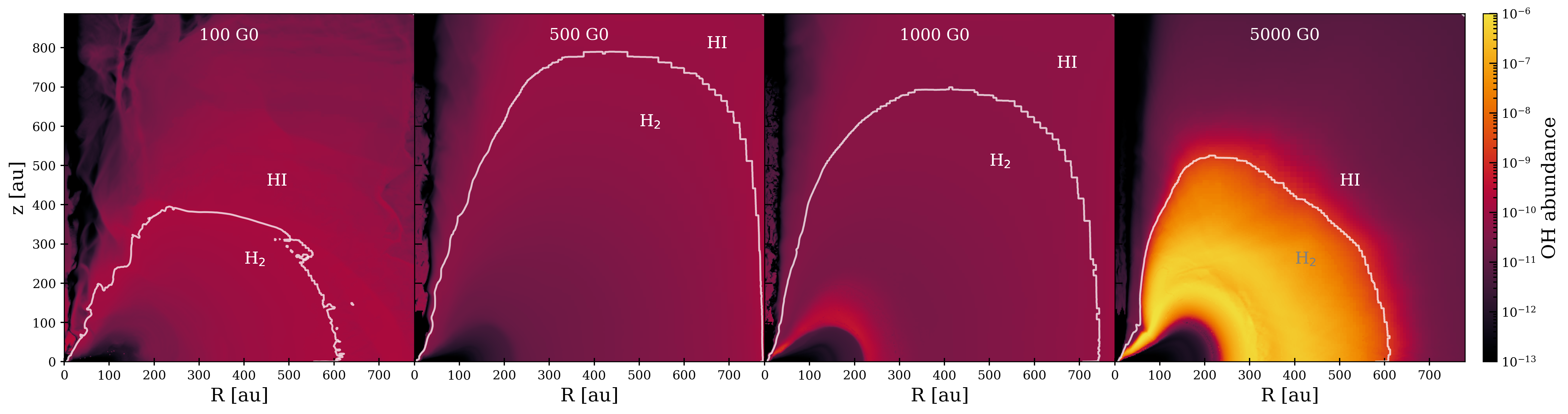}
    \includegraphics[width=\textwidth]{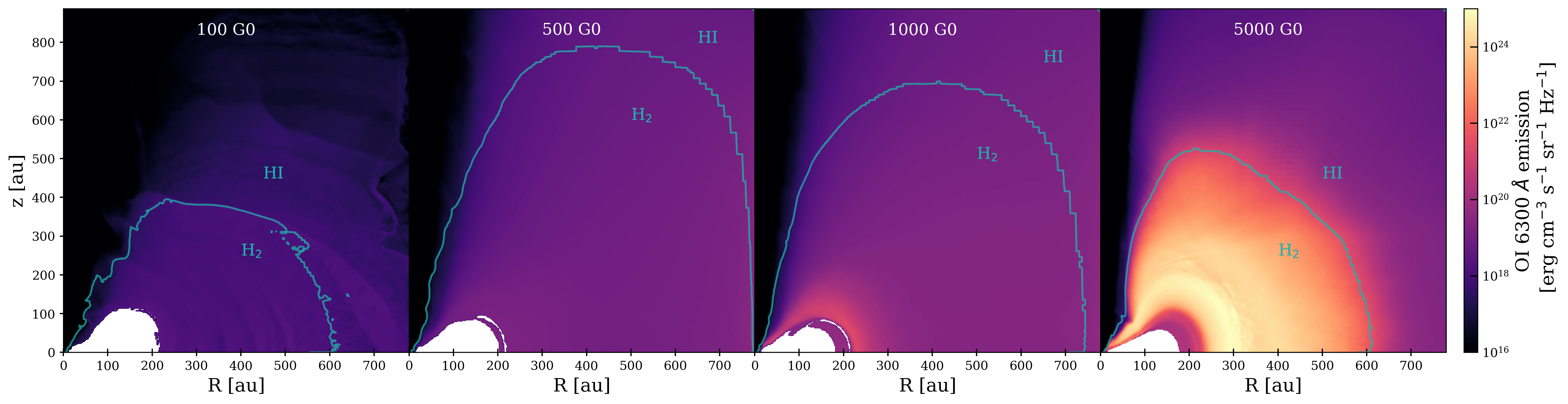}
    \caption{\emph{Top}: Density and temperature distributions of a 100~au disc around a 1~M$_\odot$ star, subject to different FUV fields. \emph{Middle}: Distribution of OH abundance in our model, for different FUV field strengths. The white contour highlights the H-H$_2$ transition. Note that the H-H$_2$ transition in the 100~G$_0$ case being at smaller radii compared to other cases due to the diffuse nature of the wind. \emph{Bottom}: Volume weighted emission map of the [O\,\textsc{i}]  6300\,\AA\ line for different FUV field strengths. Again, the teal contour highlights the H-H$_2$ transition. The white region indicates where the \oi{} is not produced due to OH reacting out by some other pathways rather than photodissociation, i.e. $R_{phot} < R_{reac}$. }
    \label{fig:2Dmodels}
\end{figure*}

There are three main factors that control the \oi{} 6300\AA\ line luminosity behaviour
\begin{enumerate}
    \item A linear scaling of the emission coefficient with the local FUV radiation field strength (equation \ref{equn:nupper}).
    \item A linear scaling of the emission coefficient with the OH abundance.
    \item Suppression of \oi{} 6300\AA\ emission due to OH reacting out by other pathways.
\end{enumerate}
The sharp increase in luminosity above a few thousand G$_0$ is contributed to by all three. However, the key factor is that the quantity $n_{\rm OH} F_{\rm FUV}$ is largest for when the density and temperature are $n\sim10^5-10^7$\,cm$^{-3}$ and $>200$\,K respectively. A density of $n\sim10^5-10^7$\,cm$^{-3}$ at lower temperatures has OH removed by chemical reactions, as does temperatures $>200\,$K and higher densities than $n\sim10^7$\,cm$^{-3}$. Figure \ref{fig:EmissionMeasure} illustrates the product of the OH number density (cm$^{-3}$) and local FUV field strength (G$_0$) as a function of temperature in two of the 1D models. Both models are a 0.6\,M$_\odot$ star, with $\Sigma_{1au}=10$\,g\,cm$^{-2}$ and a disc radius of 80\,au. The large triangles are a $10^3$\,G$_0$ environment and the circles a $10^4$\,G$_0$ environment (note that the $F_{\rm FUV}$ in value is the attenuated value of that environmental FUV field at any given point). In both cases the densest (yellow) material is near the base of the flow. Being cold and dense, OH simply reacts out quickly into other species.
Proceeding further out into the flow it becomes warmer, slightly less dense, and the FUV radiation field strength increases.  The key distinction between the $10^3$ and $10^4$\,G$_0$ cases is the steep rise in emission above T$\sim200$ K because the latter results in a warmer, denser more strongly irradiated flow. Further out still in the wind, the flow density drops to lower values where the emission is simply weaker due to the density being lower.


Our discussion so far has focused on 1D external photoevaporation simulations, which obviously are of limited geometrical accuracy. Higher dimensional models are possible but are computationally expensive. To gauge the expected implications of using a 1D geometry in our analysis we compared with 4 new 2D-axisymmetric models, similar to those of \cite{2019MNRAS.485.3895H}. These are irradiated by an isotropic bath of FUV radiation. Our new calculations are at higher resolution than those of \cite{2019MNRAS.485.3895H} and consider a wider range of UV fields. The density and temperature structure of these models is given in the upper two sets of panels in Figure \ref{fig:2Dmodels}. The third panels of Figure  \ref{fig:2Dmodels} show the OH abundance and contours of the H-H$_2$ transition. The lower panels show the volume weighted [OI] 6300\AA\ emission map.

Above 500\,G$_0$, as the UV field increases the H-H$_2$ transition pushes closer to the disc and the flow gets denser and warmer. We integrated the emission using the axisymmetry of the simulation and find that the [OI] 6300\AA\ emission is typically one order of magnitude higher than the estimates from analogous 1D models. We hence proceed with the expectation that the 1D models that we are focusing on provide conservative line luminosity estimates.


\subsection{The effect of PAH abundance on the line luminosity}
Delocalised electrons in PAHs are easily liberated, making them potentially the most significant contributor to the heating rate in PDRs and hence possibly the most important heating mechanism for external photoevaporation \citep[][]{2016MNRAS.457.3593F}. Their abundance plays a key role in setting the flow density, temperature and ultimately the mass loss rate (Haworth et al. in preparation). Given the sensitivity of [OI] 6300\AA\ to the density and temperature discussed in \ref{sec:scaling} that PAH abundance is therefore also likely to affect that line luminosity.

Considering a 100~au disc model around a 1~M$_\odot$ star, we look at the effects of varying the PAH abundance on the \oi{} line luminosity. The parameter $f_{\textrm{PAH}}$ within \textsc{torus-3dpdr} external photoevaporation calculations regulates the PAH-to-dust ratio relative to the ISM, meaning that $f_{\textrm{PAH}} = 0.1$ corresponds to 10 per cent of the canonical PAH-to-dust ratio in the ISM. Note that the dust-to-gas ratio in the disc can also be (and is, in external photoevaporation calculations) depleted separately, which will further act to reduce the PAH-to-gas ratio. The canonical FRIED grid abundance is about 1/330 of the ISM PAH-to-gas ratio ($f_{PAH}=0.1$, dust to gas mass ratio of $3\times10^{-4}$). \cite{2013ApJ...765L..38V} measured a depletion of the PAH-to-gas ratio in the proplyd HST 10 (OW94 182-413) of a factor 50, so the FRIED calculations are thought to be conservative.

\begin{figure}
    \centering
    \includegraphics[width=\columnwidth]{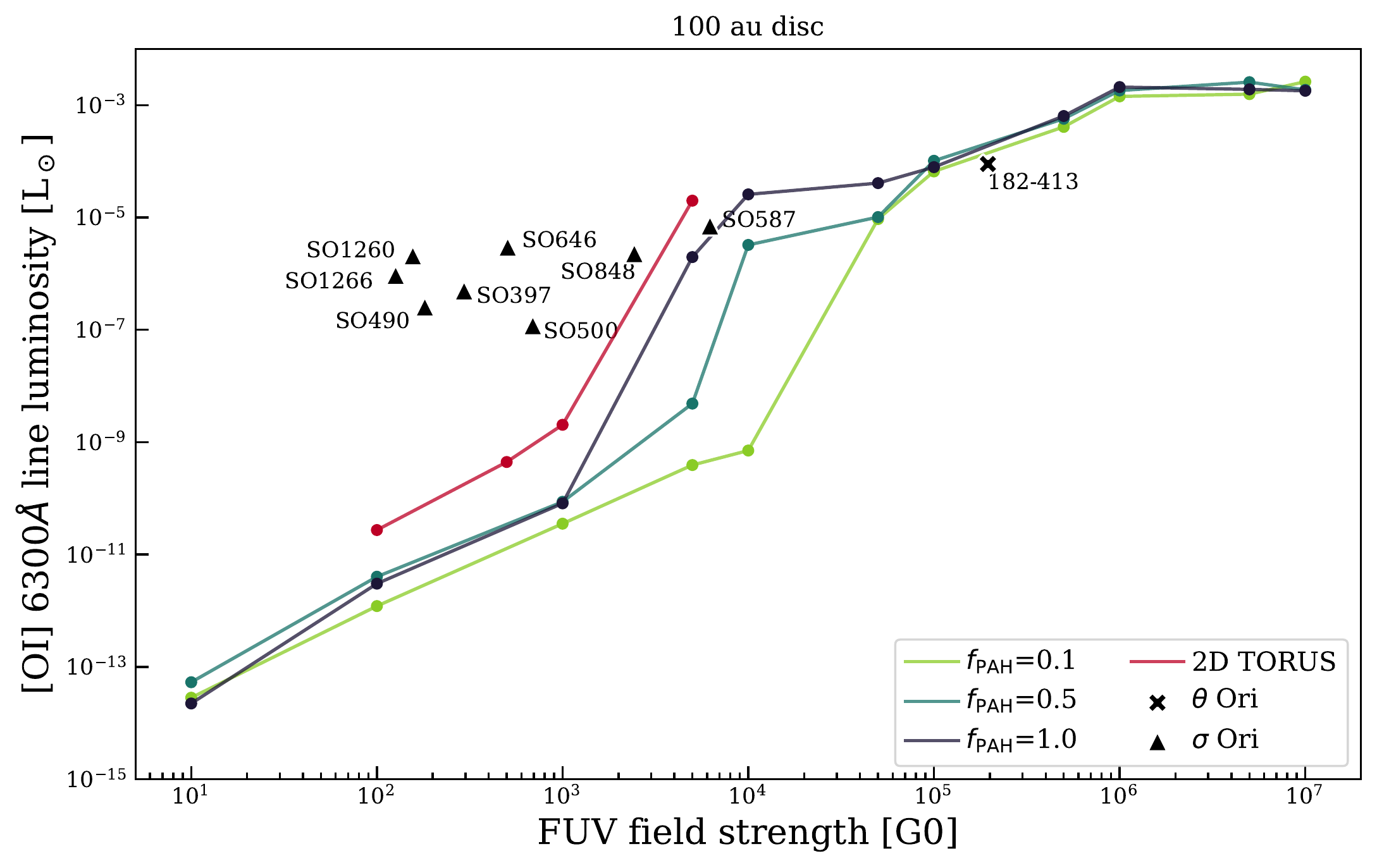}
    \caption{\oi{} 6300~\AA\,total luminosity of discs as a function of the FUV field strength. Green points and lines are the 1D FRIED models for a 100~au disc with different PAH fractions. Black triangles represent the observed \oi{} 6300~\AA\,line luminosities and FUV field that the disc would be exposed to if its true separation from the closest O star were the same as the projected separation \protect\citep{2014A&A...569A...5N}. The data include the candidate evaporating system SO587 in $\sigma$ Ori \protect\citep{2009A&A...495L..13R} and the ONC proplyd 182-413 \protect\citep{1998ApJ...502L..71S, 1998AJ....116..293B}.}
    \label{fig:OIlum_G0}
\end{figure}

 Figure~\ref{fig:OIlum_G0} again shows the predicted \oi{} 6300\AA~line luminosity as a function of the ambient UV field strength, but for different values of $f_{PAH}$. At UV field strengths in the range $10^3-10^5$G$_0$, the reduced heating at low $f_{PAH}$ values results in lower mass loss rates and a cooler wind, whereas in Figure \ref{fig:EmissionMeasure} we saw that dense, warm flows are required for strong [OI] emission. The \oi{} 6300\AA~line luminosity is therefore lower over that FUV field strength range for lower $f_{PAH}$. Note that at very high UV field strengths $>10^{5}$G$_0$ the UV field is sufficiently high that the behaviour is independent of the PAH abundance.

 The final point worth noting is that if SO587 really is externally photoevaporating \citep{2009A&A...495L..13R} $f_{PAH}>0.5$ are required to be compatible with that interpretation. Note that this is plausible though given that for an $f_{PAH}$ of 0.1 the PAH-to-gas ratio is over a factor 6 lower than that inferred for 182-413 by \cite{2013ApJ...765L..38V}, as discussed above.



\section{Comparing with ONC proplyd luminosities}
\label{sec:ONCcomparison}
Before applying our theoretical results to problems such as identifying external photoevaporation in distant regions or weak UV fields, we compare with the [OI] 6300\AA\ luminosities of known proplyds in the ONC. Making this comparison is somewhat challenging, in part because very few values are quoted in the literature \citep[e.g. 182-413][]{1998ApJ...502L..71S,1998AJ....116..293B} and also because the line can originate from both the PDR (included in our models) and the ionisation front (not included in our models).

\begin{figure*}
  \centering
  \includegraphics[width=\linewidth]{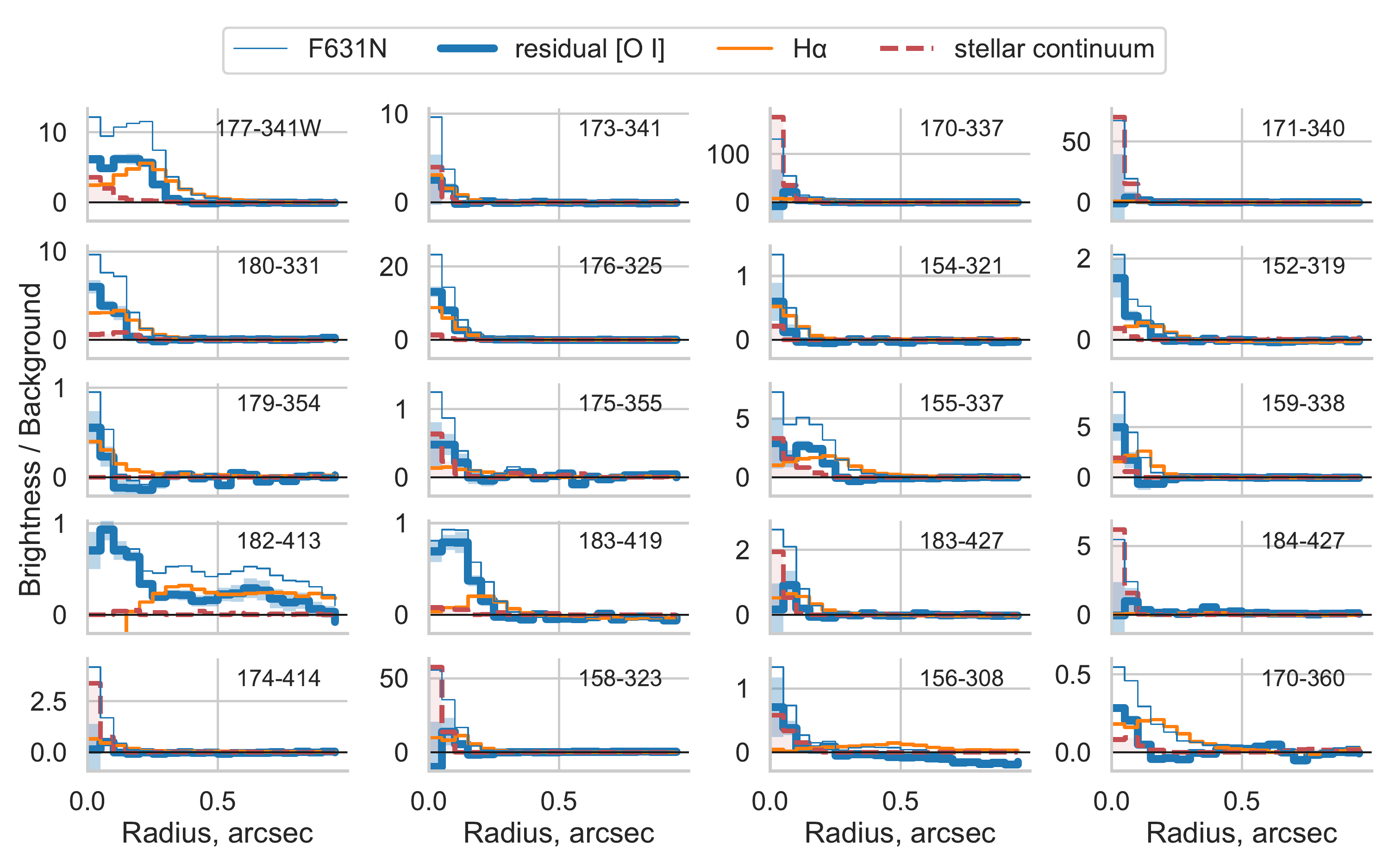}
  \caption{
    \oi{} PDR emission from HST imaging of Orion proplyds.
    Thin blue lines show the brightness profiles in the
    F631N filter averaged over the hemisphere of the proplyd
    that faces the ionizing star.
    Heavy blue lines show the residual \oi{} emission after
    subtracting the contributions from the ionization front
    (modeled by scaling the \ha{} profile, orange line)
    and the proplyd's central star (red-dashed line).
    Blue shading indicates uncertainties in the residual profile.
  }
  \label{fig:proplyd-oi-profile}
\end{figure*}

We therefore combined archival HST F631N (\oi{} 6300\AA\, and continuum), F547M (continuum) and F656N (\ha) data \citep[O'Dell, private comm., but see e.g.][for additional information]{1998AJ....115..263O, 2001ARA&A..39...99O} of 22 proplyds in the ONC to distinguish the different components. We describe here the data reduction that lead to recovering additional new information on the [OI] emission of previously known proplyds.
By assuming that the \oi/\ha{} ratio in the ionization front
is the same as in the background nebula,
we are able to use a scaled version of the \ha{} image to
fit and remove the contribution of the ionization front
to the \oi{} emission.
The continuum F547M filter is used to estimate the contribution
of the proplyd's central star to the F631N filter,
which is also subtracted to leave a residual \oi{} profile,
which should primarily arise in the neutral evaporation flow
from the proplyd disc.
Resulting profiles are shown in Figure~\ref{fig:proplyd-oi-profile},
while the PDR \oi{} 6300\AA\ luminosity components for the proplyds are summarised in Table~\ref{tab:ONC_LOI}.

\begin{table*}
    \centering
    \begin{tabular}{ccccccccc}
    \hline
    Proplyd & Projected $\theta^1$C  & PDR \oi{} 6300\AA\  & Estimated External & Estimated Disc  \\
    Name & Separation [\arcsec] &  Luminosity [$10^{-4}L_\odot$] & FUV [$10^6$G$_0$] & Radius [au]  \\
    \hline
    177-341W & 25.433   & $1.466^{+0.194}_{-0.195}$ & 0.96 & 56.3 & \\
    \\[-1em]
    173-341 & 22.345   & $0.0875^{+0.0566}_{-0.0528}$ & 1.24 & 17.7 & \\
    \\[-1em]
    170-337 & 15.998  & $1.307_{-1.133}^{+1.623}$  & 2.43 & 40.4 & \\
    \\[-1em]
    171-340 & 18.873   & $0.2923_{-0.2207}^{+0.9277}$ & 1.74 & 39.2 & \\
    \\[-1em]
    180-331 & 24.697   & $0.3079_{-0.0656}^{+0.0717}$  & 1.02 & 26.6 & \\
    \\[-1em]
    176-325 & 16.343   & $0.6304_{-0.0802}^{+0.0802}$  & 2.33 & 17.6 & \\
    \\[-1em]
    161-328 & 7.564   & $0.0151_{-0.0086}^{+0.0191}$    & 10.8 & 10.0 & \\
    \\[-1em]
    158-327 & 10.711  & $0.1867_{-0.1335}^{+0.4129}$   & 5.42 & 40.9 & \\
    \\[-1em]
	154-321 & 16.701  & $0.0478_{-0.0323}^{+0.0323}$  & 2.23 & 13.5 & \\
    \\[-1em]
    152-319 & 19.243  & $0.1661_{-0.0483}^{+0.0488}$   & 1.68 & 22.3 & \\
    \\[-1em]
    179-354 & 37.697   & $0.0218_{-0.0090}^{+0.0090}$  &  0.44 & 15.9 & \\
    \\[-1em]
    175-355	 & 35.724   & $0.0311_{-0.0129}^{+0.0129}$ &  0.49 & 20.0 & \\
    \\[-1em]
    155-337 & 20.310   & $1.110_{-0.2973}^{+0.3180}$   & 1.51 & 45.1 & \\
    \\[-1em]
    159-338 & 17.298   & $0.3035_{-0.1129}^{+0.1129}$  &  2.08 & 15.1 & \\
    \\[-1em]
    182-413 & 56.782   & $0.6795_{-0.2425}^{+0.2478}$  &  0.19 & 109.1 & \\
    \\[-1em]
    183-419 & 62.198   & $0.1209_{-0.0193}^{+0.0226}$   & 0.16 & 39.2 & \\
    \\[-1em]
    183-427 & 69.641   & $0.0589_{-0.0315}^{+0.0427}$   &  0.13 & 30.4 & \\
    \\[-1em]
    184-427 & 69.706  & $0.1580_{-0.1167}^{+0.1960}$   & 0.13 & 87.6 & \\
    \\[-1em]
    174-414 & 52.634   & $0.0270_{-0.0114}^{+0.0313}$ &  0.22 & 25.8 & \\
    \\[-1em]
    158-323 & 9.476   & $3.186_{-2.063}^{+3.231}$   & 6.92 & 35.8 & \\
    \\[-1em]
    156-308 & 19.462   & $0.1093_{-0.0566}^{+0.0605}$  & 1.64 & 25.7 & \\
    \\[-1em]
    170-360 & 37.327   & $0.0226_{-0.0038}^{+0.0038}$   & 0.45 & 21.3 & \\
    \\[-1em]
    \hline
    \end{tabular}
    \caption{The [OI] 6300\AA\ line luminosity in the PDR component of ONC proplyds. We also include the projected separation from $\theta^1$C and the corresponding FUV field incident upon the proplyds assuming the projected separation is the true separation and geometric dilution of the FUV radiation field.  }
    \label{tab:ONC_LOI}
\end{table*}

\begin{figure}
    \centering
    \includegraphics[width=\columnwidth]{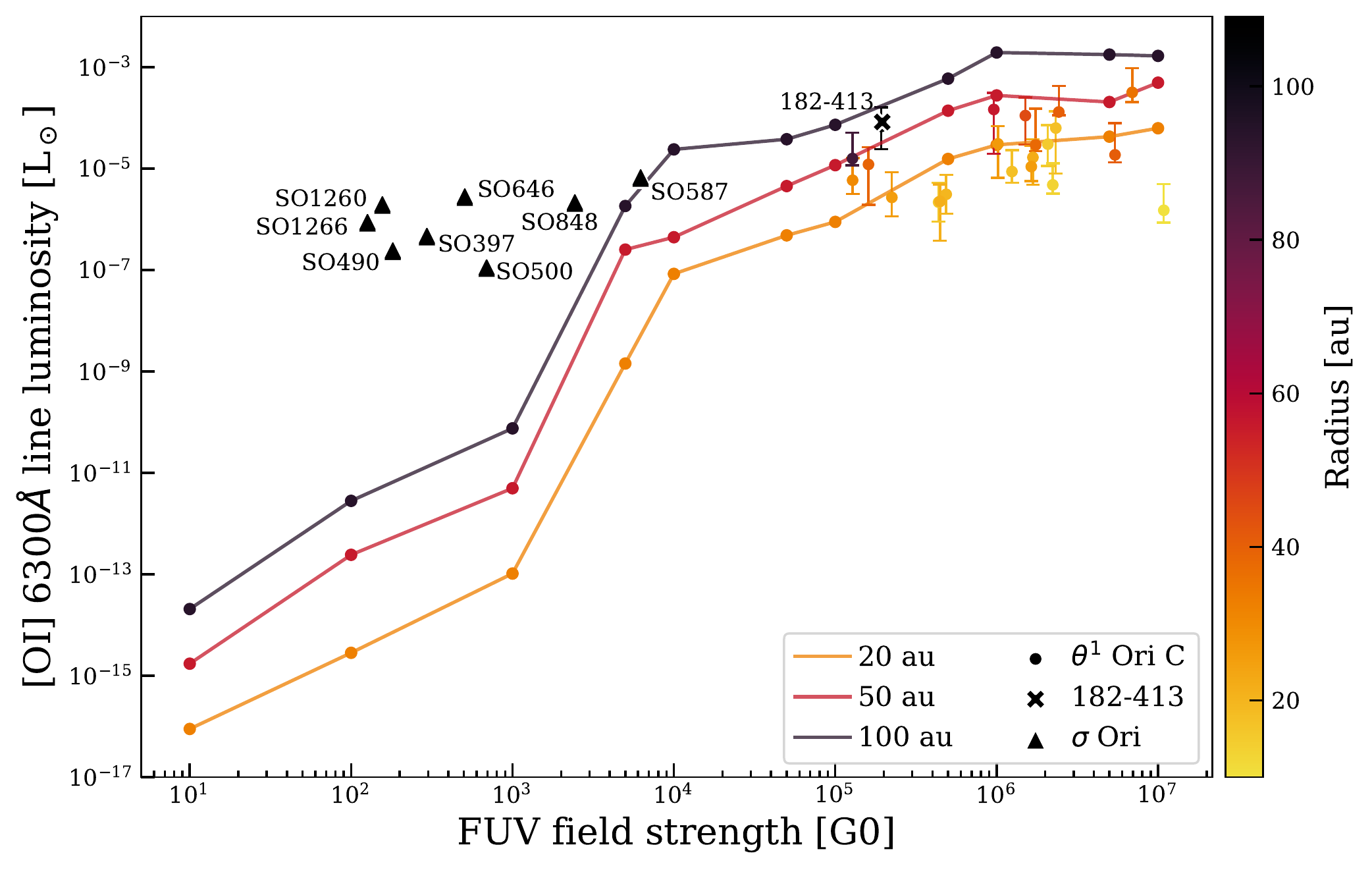}
    \includegraphics[width=\columnwidth]{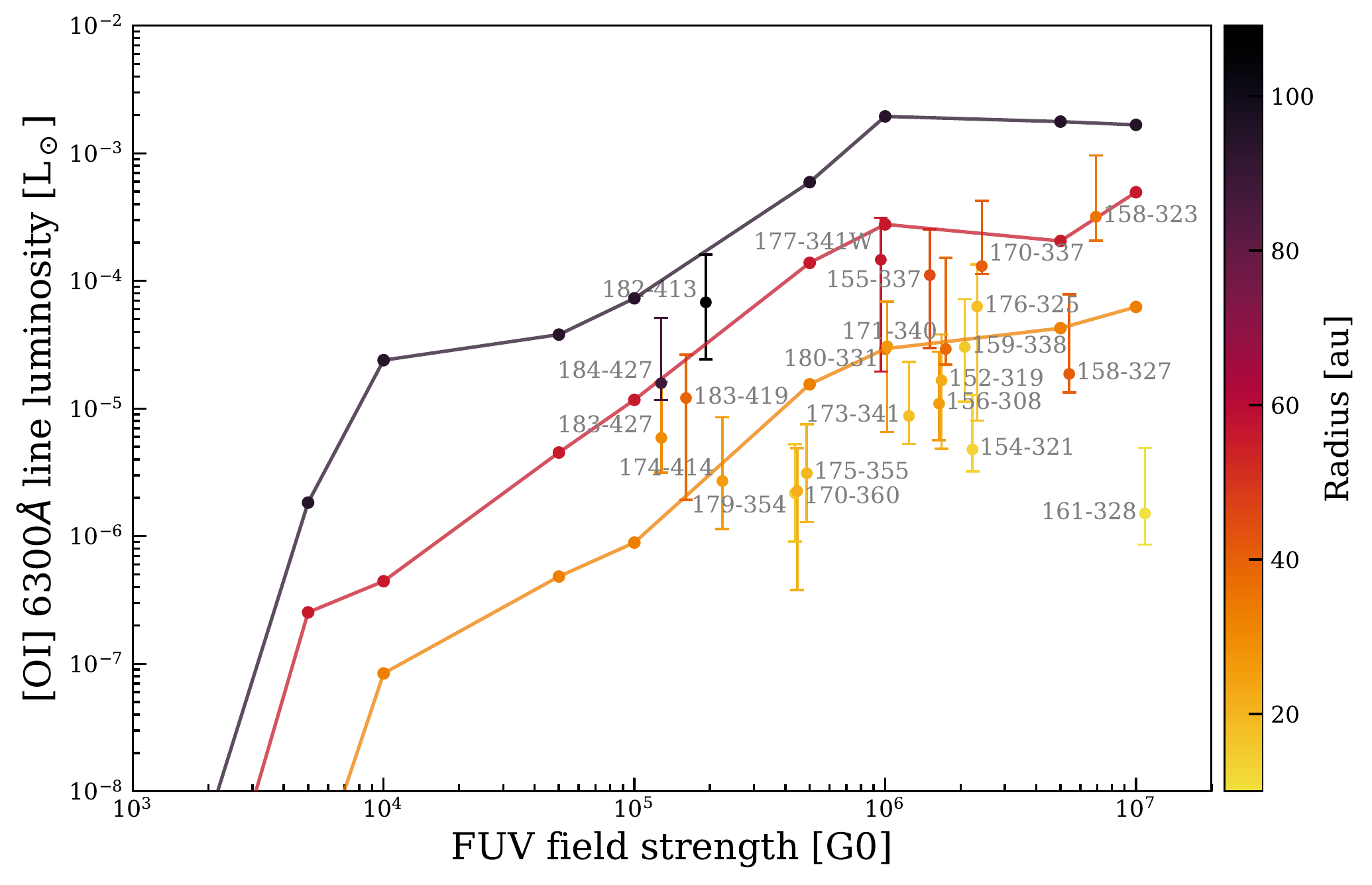}
    \caption{Model and observed [OI] 6300\AA\ line luminosities. The lower panel is the same, but with focusing on the dynamic range spanned by the observations. In each panel the triangles are $\sigma$ ori data from \protect\cite{2014A&A...569A...5N}. The other points are our estimates for the PDR in ONC proplyds. The models are for discs of 100, 50 and 20au. With a few exceptions, the ONC proplyds have higher [OI] 6300\AA\ luminosities than the discs in $\sigma$ ori, consistent with our models for $\sim20-50$\,au discs.  }
    \label{fig:ONCComparison_LOI}
\end{figure}
Figure \ref{fig:ONCComparison_LOI} shows the proplyds and $\sigma$ Ori line luminosities as a function of FUV field strength, compared with our model estimates for discs with radii of 20, 50 and 100\,au. The proplyd data points are colour coded by the approximate radius of [OI] emission, which is an upper limit to the disc radius.
We are not trying to model specific systems here, but note that the models predictions are broadly consistent with observations for disc radii $\leq50$\,au. Our radius estimates are approximate upper limits, but we also note that \cite{2020ApJ...894...74B} found CO radii that are typically around 50\,au in the ONC at distances larger than the proplyds we consider here, so again consistent with the small radii expected from our models.

Most of the proplyds have higher [OI] 6300\AA\ luminosities than the $\sigma$ Ori discs (typically by an order of magnitude). The $\sigma$ Ori discs have stronger [OI] 6300\AA\ luminosities than our model at low UV field strengths because the origin of the line in those systems is the internal wind, which we do not include in our model. We note that SO587 had been proposed as an externally photoevaporating disc based on its relatively high [OI] luminosity \citep{2009A&A...495L..13R}. Our more extended disc models ($\sim$100~au) are compatible with this interpretation.

Overall our models predict that in high UV environments the [OI] 6300\AA\ luminosities are substantially higher than those typically seen from internal photoevaporation alone and this is supported by our measured values for the ONC proplyds.

\section{Scaling of the \boldmath\oi{} luminosity with accretion luminosity in different UV environments}
\label{sec:LOIvsLacc}

\begin{figure*}
    \centering
    \includegraphics[width=0.49\textwidth]{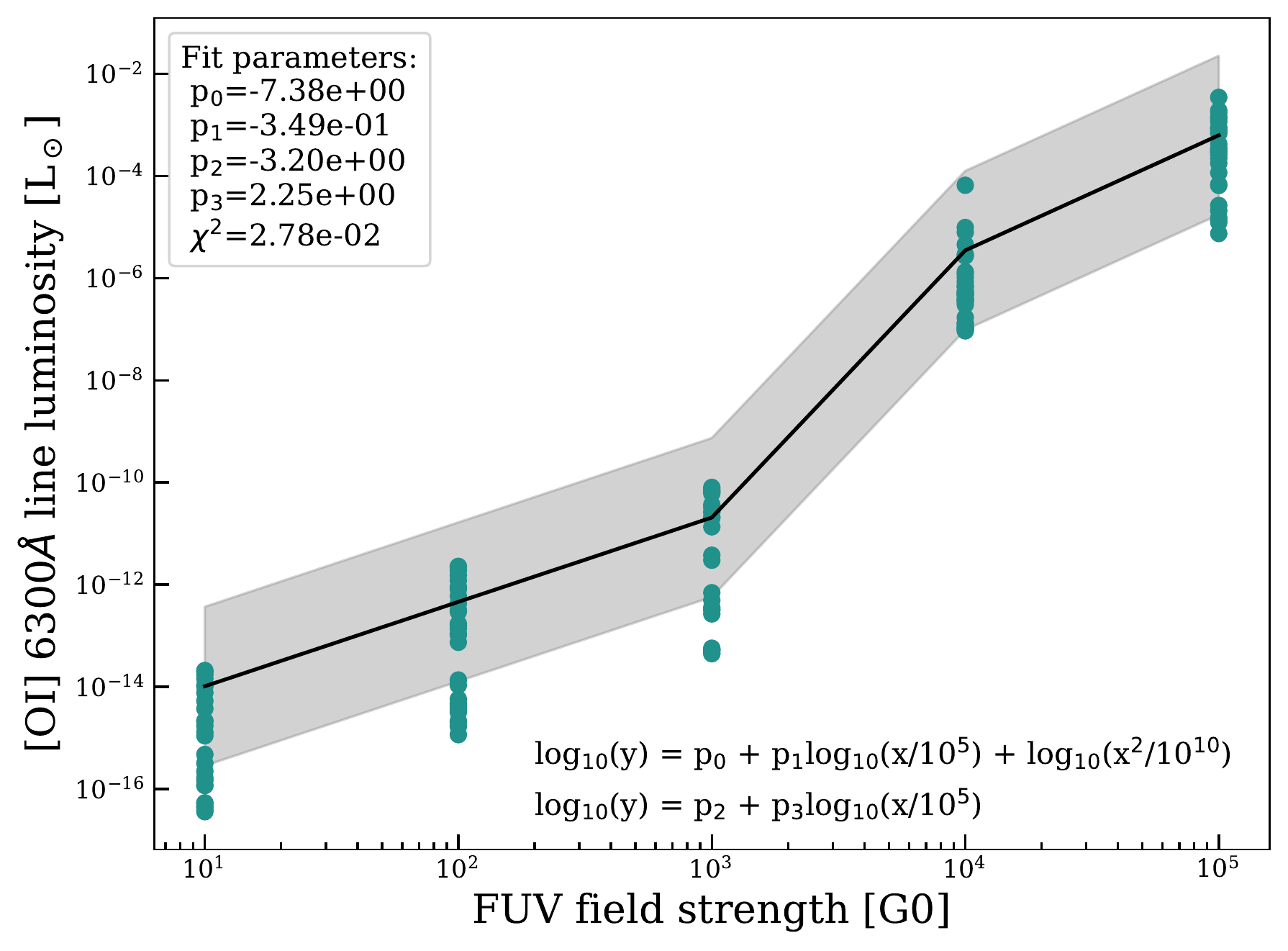}
    \includegraphics[width=0.49\textwidth]{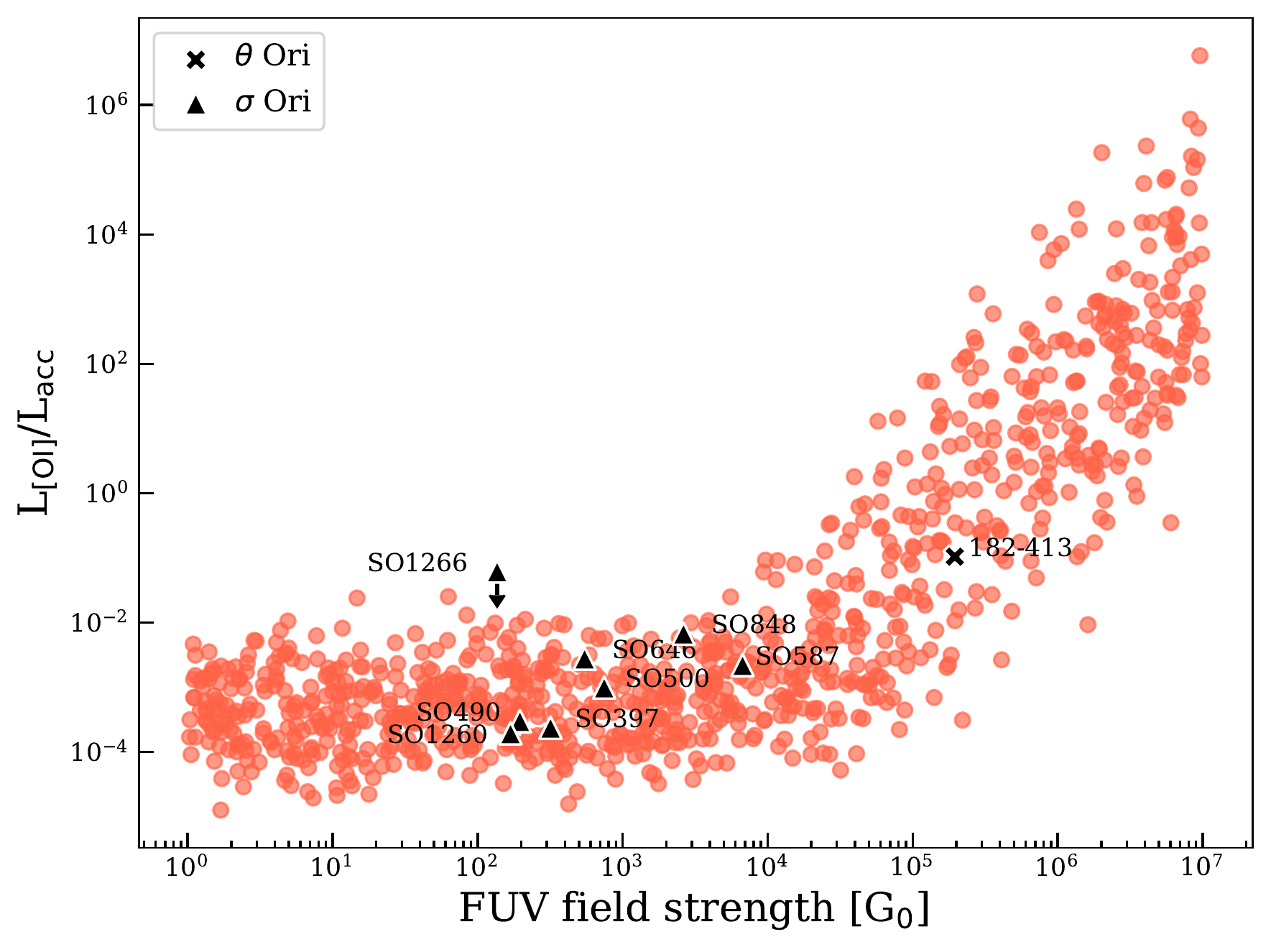}
    \caption{\emph{Left}: Predicted 1D model \oi{} 6300$\AA$ line luminosities. The spread at each value of the FUV field strength stems from variations in the disc mass/radius. The black line is the fit to the models represented by the equation in implicit form. The error on the fit, calculated as the $\chi^2$, is highlighted by the grey band.
    \emph{Right}: Modelled ratio of the total \oi{} 6300\AA\ LVC line luminosity to accretion luminosity. Up to around $10^4$\,G$_0$ the \oi{} line is dominated by inner emission, which correlates with the accretion luminosity \protect\citep[the distribution here is based on][]{2018A&A...609A..87N}. Above around $10^4$\,G$_0$ the \oi{} contribution from external photoevaporation becomes significant, which does not correlate (or at least does not correlate so strongly) with the accretion luminosity. The black triangles indicate data in $\sigma$ Ori taken from the literature \protect\citep[][]{2014A&A...569A...5N, 2012A&A...548A..56R}. The ONC proplyds in which we have detected \oi{} emission do not have accretion luminosity estimates, however for illustrative purposes we show the location of 182-413 assuming an accretion luminosity from the \protect\cite{2018A&A...609A..87N} scalings for the stellar mass of the object. }
    \label{fig:modelDistribution}
\end{figure*}

Our models predict a strong increase in the \oi{} 6300\AA\ line luminosity as a function of the FUV field strength. However, there is also a known correlation between the \oi{} 6300\AA\ line luminosity and accretion luminosity \citep[][]{2013ApJ...772...60R, 2014A&A...569A...5N, 2018A&A...609A..87N}. Using numerical simulations, \cite{2016MNRAS.460.3472E} showed that such correlation could be explained by the correlation between the line luminosity and the EUV flux, which is dominated by accretion photons. A larger portion of wind is heated by higher accretion luminosities, enhancing the collisionally excited [OI] emission. In an X-Shooter survey of T Tauri stars in Lupus, Chameleon and $\sigma$ Orionis \cite{2018A&A...609A..87N} find
\begin{equation}
    \log_{10}\left(\frac{L_{\textrm{[OI]}}}{L_\odot}\right) = 0.6\log_{10}\left(\frac{L_{\textrm{acc}}}{L_\odot}\right) -4.1
    \label{equn:LOI_Lacc}
\end{equation}
for the low velocity component of the line, with a spread of about 1.6 dex.
However for external photoevaporation only (ignoring the internal winds) there should be no strong direct correlation between the accretion luminosity and the [OI] 6300\AA~line luminosity since the heating is dominated by the external radiation field. This could provide an important diagnostic for external photoevaporation in distant high UV environments. This is subject to some possible minor caveats, in that there is some suggestion that external photoevaporation might have an effect on mass accretion rates. i) For a given disc mass and age, there is some evidence that mass accretion rates are higher in star forming regions with higher UV environments \citep[see][for a comparison between discs in Lupus and in Orion]{2017MNRAS.468.1631R}. ii) External photovaporation preferentially depletes the outer disc, down to a critical radius. Therefore, initially the gas surface density is depleted and truncated over time from the outside-in. Once the critical radius is reached, the external mass loss is not as effective anymore and thus viscous spreading of the disc takes over (assuming the disc is viscously evolving) reducing the amount of gas that can flow into and replenish the inner disc before ultimately accreting onto the star. This would then lead to slightly lower mass accretion rates onto the star, mainly due to lower gas surface density \citep[][]{2022MNRAS.514.2315C, 2022MNRAS.515.4287W}. Neither of these potential caveats would be strong enough to dominate over the well studied scaling of the internal [OI] and accretion luminosities.

To make a first exploration of the viability of the \oi{} 6300\AA\ line as a diagnostic of external photoevaporation in distant high UV environments, we randomly sample accretion luminosities in different UV environments based on the models here and the observations of \cite{2014A&A...569A...5N}. We assume that the accretion luminosities are distributed as a Gaussian of the form
\begin{equation}
    P(L_{\textrm{acc}}) \propto \exp\left[-\frac{1}{2}(\log_{10}(L_{\textrm{acc}})-2)^2\right]
\end{equation}
i.e. a Gaussian with standard deviation in $\log_{10}(L_{\textrm{acc}})$ of 1 centred on $\log_{10}(L_{\textrm{acc}})=-2$. That randomly sampled accretion luminosity then corresponds to a \oi{} 6300\AA\ luminosity following equation \ref{equn:LOI_Lacc}, where we include a random modification to the logarithmic \oi{} luminosities of $\pm 0.8$ to account for the spread in the \cite{2018A&A...609A..87N} data. For each sampled point we then select a random ambient FUV radiation field strength from $1-10^6$\,G$_0$ and introduce an additional \oi{} luminosity component based on our models of external photoevaporation in different environments. Specifically we approximate the models varying with FUV field strength as
\begin{equation}
    \log_{10}\left(\frac{L_{[\textrm{OI}]}}{L_\odot}\right) =
\begin{cases}
    -7.4 - 0.35\log_{10}\left(\frac{F_{\textrm{FUV}}}{10^5\,\textrm{G}_0} \right) + \\
    \qquad \quad + \log_{10}\left(\left[\frac{F_{\textrm{FUV}}}{10^5\,\textrm{G}_0}\right]^2 \right), & \text{if } F_{\textrm{FUV}} \leq 10^3\,\textrm{G}_0 \\
    -3.2 + 2.25\log_{10}\left(\frac{F_{\textrm{FUV}}}{10^5\,\textrm{G}_0} \right), & \text{if } F_{\textrm{FUV}} > 10^3\,\textrm{G}_0 \\
    & \quad F_{\textrm{FUV}} \leq 10^4\,\textrm{G}_0 \\
    -0.25 + 5.2\log_{10}\left(\frac{F_{\textrm{FUV}}}{10^5\,\textrm{G}_0} \right), & \text{if } F_{\textrm{FUV}} > 10^4\,\textrm{G}_0
\end{cases}
\end{equation}
with a random spread of $\pm2$ dex. The fit has been performed using the \verb|mpfit| python module. The spread corresponds to the error on the fit calculated as the $\chi^2$.
The resulting ratio of \oi{} luminosities to accretion luminosities for a distribution of 1000 randomly sampled discs is shown in Figure \ref{fig:modelDistribution}. Below $10^4$\,G$_0$ the \oi{} luminosity is dominated by emission from the inner disc that scales with the accretion luminosity following \cite{2018A&A...609A..87N}. At stronger FUV fields, however, there is a significant rise in \oi{}-to-accretion luminosities owing to the additional external component.

Accretion rate estimates exist for the Sigma Orionis discs by \cite{2014A&A...569A...5N}, however the ONC proplyds for which we measure the \oi{} luminosity do not yet have accretion rate estimates \citep[the proplyd accretion rate estimates from][do not overlap with our sample]{2012ApJ...755..154M}. We are hence limited in our ability to test this prediction on local sources (though we reiterate that in \ref{sec:ONCcomparison} we have demonstrated that the prediction of high \oi{} luminosity is borne out in ONC proplyds).

Nonetheless, the proplyd 182-413 has a strong [OI] 6300\AA\ line luminosity that is spatially resolved and clearly associated with the external photoevaporative wind, in a relatively strong UV environment. This would make it an ideal test of the line-to-accretion luminosity diagnostic. Unfortunately there are no accretion rate measurements for the object.  To enable us to make a basic, approximate estimate of this ratio we place constraints on the stellar mass and consider the range of plausible accretion rates for a star of such a mass. The stellar mass is estimated using the K-band magnitude and J-K colour, coupled with UKIDSS magnitudes predicted from \cite{2015A&A...577A..42B}. 182-413 has a K band magnitude of 13.31 and a J-K colour of 2, which for a reddening factor $R_V=5.5$ \citep{2021ApJ...908...49F} results in a K band magnitude of approximately 12.31  \citep[see Figure 2 of][for the reddening vector]{2022MNRAS.512.2594H}. The K band magnitude is a strong function of stellar mass at low stellar masses,  meaning it can be used to make approximate constraints on the stellar mass as being 0.18-0.28\,M$_\odot$. We then use the correlations in \cite{2018A&A...609A..87N} between \oi{} luminosity and both stellar mass and accretion luminosity, to find an rough estimate of the accretion luminosity for 182-413. Figure~\ref{fig:modelDistribution} shows the line-to-accretion luminosity ratio for 182-413, which lies well within the spread of our predictions.

The power that this diagnostic would have over simply searching for \oi{} luminosities that are unusually high is that it unambiguously distinguishes bright internal contributions associated with high accretion and external contributions to the line luminosity. The exciting potential utility of this new diagnostic would be the ability to identify external photoevaporation in distant high UV environments where proplyds cannot be spatially resolved.

\section{Discussion}

\subsection{Impact of EUV radiation}
The external photoevaporative mass loss rate from sub-critical discs (disc radius less than the gravitational radius) is dominated by FUV radiation. Although it doesn't alter the mass loss rate, ionising EUV photons establish an ionisation front downstream in the wind which affects the morphology and the observable characteristics near to the ionisation front. Our calculations are FUV-only, but we gauge the approximate location of the ionisation front assuming ionisation equilibrium and exclude [OI] emission beyond that point.

The disc is irradiated by an ionising flux $n_{i,0}$\,cm$^{-2}$\,s$^{-1}$. The number of ionising photons required to keep a cell of width $\Delta R$ with $n_H$ hydrogen nuclei ionised is
\begin{equation}
    n_i = n_H^2 \alpha_B\Delta R
    \label{equn:nioreq}
\end{equation}
where $\alpha_B$ is the case B hydrogen recombination coefficient ($\alpha_B \approx 2.7\times10^{-13}$cm$^3$s$^{-1}$). We integrate equation \ref{equn:nioreq} on a cell-by-cell basis until the sum of ionising photons required to maintain ionisation equals $n_{i,0}$. If we remove all the emission in the portion of the grid that would be ionised we find that the EUV has a negligible impact on the [OI] emission from our models.

\subsection{The possible utility of \boldmath\oi{} 6300\AA\ for diagnosing external photoevaporation in unresolved systems}



A key implication of Figures \ref{fig:OIlum_G0}, \ref{fig:ONCComparison_LOI} and \ref{fig:modelDistribution} is that the \oi{} brightness increases above typical levels for internal disc winds above $\sim10^4$~G$_0$. Our $10^5$~G$_0$ environment models are sufficiently bright that the level of sensitivity in \cite{2014A&A...569A...5N} \citep[who used data from][]{2011AN....332..242A, 2012A&A...548A..56R} would in principle be sufficient to detect \oi{} from them in distant star forming regions like Carina \citep[which is at 2.3kpc,][]{2006MNRAS.367..763S}. For example, the faintest source in \cite{2014A&A...569A...5N} detected with X-Shooter has a flux of  $\sim3\times10^{-17}$erg\,cm$^{-2}$\,s$^{-1}$. A typical disc in our model set in a $10^5$~G$_0$ field at 2.3\,kpc is predicted to have a \oi{} flux of around $3\times10^{-15}$erg\,cm$^{-2}$\,s$^{-1}$, so should be detectable. However there may be significant technical challenges associated with the subtraction of the background nebula and overcrowding. If those can be circumvented, bright \oi{} lines (brighter than the threshold expected for the flux to be dominated by the disc or an inner wind) associated with YSOs could provide evidence for external photoevaporation in distant star forming regions and could possibly be achieved with X-Shooter.

\subsection{Directly separating the \boldmath\oi{} 6300\AA\ emission contributions from internal and external winds }
\begin{figure}
    \centering
    \includegraphics[width=\columnwidth]{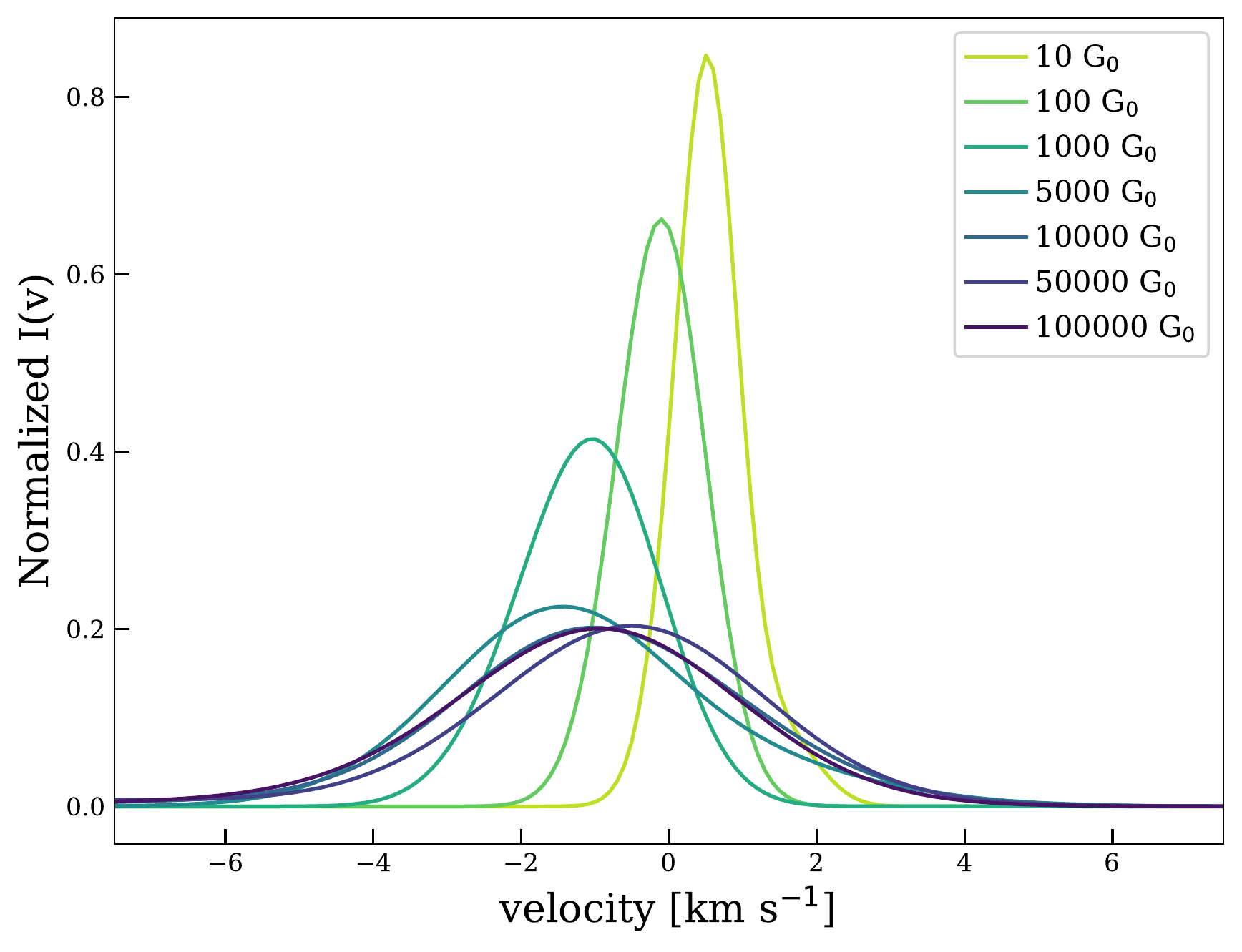}
    \caption{Theoretical line profiles for the \oi{} 6300\AA\ emission line, calculated from our 1D models at different FUV field strengths. Each line has been normalised to the total integral of the line. Note that this is the radial component of the velocity (i.e. not including the azimuthal), which is typically low compared to the velocity of internal winds.}
    \label{fig:lineProfiles}
\end{figure}
In Section \ref{sec:LOIvsLacc} we considered the total \oi{} 6300\AA\ line luminosity from both internal and external photoevaporative winds. Here we briefly discuss distinguishing those internal and external line contributions directly in observations of any given system.

The PDR components of external photoevaporative winds are relatively slow, with temperatures of hundreds to a few thousand Kelvin and flow velocity of around a few km\,s$^{-1}$. These external winds are predominantly launched from the disc outer edge, although there are components that are launched from the disc surface layers towards the outer part of the disc \citep{2019MNRAS.485.3895H}. This results in narrow line profiles, though these do broaden slightly at high FUV field strengths, as illustrated in Figure \ref{fig:lineProfiles}. On the other hand, low velocity components associated with internal photoevaporative winds have been observed to show typical blue-shifts of a few km\,s$^{-1}$, up to tens of km\,s$^{-1}$ in some cases, and widths of $\sim 25-30$\,km\,s$^{-1}$ in the narrowest cases \citep[][]{2016ApJ...831..169S, 2018ApJ...868...28F, 2019ApJ...870...76B}. Additionally, such winds are launched with a geometry that is more perpendicular (though not perfectly perpendicular) to the disc surface compared to the case of externally photoevaporated winds \citep[e.g.][]{2019ApJ...870...76B}.

The implications of the line properties and geometries of internal and external winds, combined with the fact that the internal wind is brighter in \oi{} 6300\AA\ up to $\sim10^4$\,G$_0$ is that for the majority of viewing angles we do not expect to spectrally discern the inner and outer contributions to the line flux. The external contribution will be dominated by the brighter, broader internal emission. In very high UV environments it might be possible to discern a central narrow line with broad wings. However, to spectrally distinguish internal and external winds with [OI] the most likely possibility is for near edge-on discs, since in that case the internal wind is broadly perpendicular to the line of sight, whereas the external is predominantly along the line of sight.

Overall then we generally expect that observationally distinguishing internal and external winds in \oi{} 6300\AA\ requires both spatially and spectrally resolving the disc/wind. Unfortunately though this is beyond current observational capabilities. The ANDES (once called HIRES) spectrograph on the in-development European-Extremely Large Telescope (E-ELT) offers high spectral and spatial resolution that may possibly achieve this \citep{2013arXiv1310.3163M, 2016SPIE.9908E..23M}.

Another possibility is the use of spectro-astrometry, in which the orientation of the spectral slit is adjusted on the target to constrain the orientation and launching region of the wind. For example \cite{2021ApJ...913...43W} used this to find evidence for inner MHD driven winds in two T Tauri stars, ruling out the winds as photoevaporative in origin based on the flow profile. Given the very different orientation of internal and external photoevaporative winds it is conceivable that it could be a powerful tool for spectrally distinguishing internal and external photoevaporative winds with current instrumentation. However, it is beyond the scope of this work to assess the feasibility of this here using synthetic spectro-astrometry.

\section{Summary and Conclusions}
We determine the observational characteristics of the [O\,\textsc{i}]  6300\AA\ line as a tracer of external photoevaporation. This line is also widely used as a diagnostic of internal winds \citep[e.g.][]{2014A&A...569A...5N}. It is also known to be associated with spatially resolved, spectrally unresolved emission from proplyds in high UV environments \citep{1998AJ....116..293B} and was explained theoretically in terms of OH dissociation for Orion Nebula Cluster proplyds by \cite{1998ApJ...502L..71S}. Here we greatly expand upon the calculations of \cite{1998ApJ...502L..71S} to consider a wide range of UV field strengths based on PDR-dynamical simulations of external photoevaporation. Our goal is to understand what further utility there is in the [O\,\textsc{i}] 6300\AA\ line as a diagnostic of external photoevaporation. We draw the following main conclusions from this work.  \\

1) The \oi{} 6300\AA\ line luminosity is predicted to be a strong function of the FUV radiation field strength and begins undergoing a dramatic increase above around 5000~G$_0$. The [OI] 6300\AA\ emission is a result of photodissociation of OH and depends on the product of the OH number density and FUV radiation field strength. This is maximal for warm dense flows which require strong FUV radiation fields. \\

2) The increase in \oi{} 6300\AA\ line luminosity predicted by our models is consistent with the luminosities of ONC proplyds, which are markedly brighter in the line than discs in $\sigma$ Ori. Our models are also consistent with the prior suggestion that the system SO587 in $\sigma$ Orionis is externally photoevaporating if the disc is $\sim100$\,au. \\

3) We predict that the ratio of the \oi{} 6300\AA~line luminosity to accretion luminosity can help identify external photoevaporative winds in distant clusters where proplyds cannot be resolved. This ratio is relatively flat (with some scatter) in low-intermediate UV environments. Above $\sim10^4$~G$_0$ though, this ratio is anticipated to increase substantially according to our models. Unfortunately a lack of ONC proplyd accretion rates means that this prediction is not yet tested (though we note that the [OI] luminosities themselves are consistsent with the ONC proplyds). Although this has no major impact in sites such as the ONC, it could help to address a long standing issue in that we are unable to identify external photoevaporation in distant high UV environments where cometary proplyds are difficult to resolve. \\

4) The \oi{} 6300\AA\ line luminosity and ratio relative to the accretion luminosity provides a means of identifying external photoevaporation in distant clusters. However, for any given system it is difficult to directly observe the internal and external components. This is because the low velocity of external photoevaporative winds compared to internal winds means that it is difficult to spectrally distinguish the two without spatially resolving the system. Spatially resolving the inner and outer winds (e.g. with E-ELT ANDES), or perhaps using spectro-astrometry (to be determined in future work) are required  to probe the external photoevaporative \oi{} line profile when the internal wind dominates the spectrum in spatially unresolved observations.   \\


\section*{Data Availability}
The models used in this paper will be made available upon reasonable request to the authors.

\section*{Acknowledgements}
We thank the anonymous reviewer for their comments on the manuscript. We also thank Carlo Manara, Anna McLeod, Dominika Itrich and Megan Reiter for useful discussions.

GB and TJH are funded by a Royal Society Dorothy Hodgkin Fellowship.
This work was performed using the DiRAC Data Intensive service at Leicester, operated by the University of Leicester IT Services, which forms part of the STFC DiRAC HPC Facility (www.dirac.ac.uk). The equipment was funded by BEIS capital funding via STFC capital grants ST/K000373/1 and ST/R002363/1 and STFC DiRAC Operations grant ST/R001014/1. DiRAC is part of the National e-Infrastructure.
\bibliographystyle{mnras}
\bibliography{molecular}

\appendix

\label{lastpage}

\end{document}